\documentclass{appolb}
\usepackage{enumitem}
\setlist[description]{style=nextline}
\usepackage{graphicx}

\begin{document}
\title{Model-independent Odderon results based on TOTEM data  on elastic proton-proton scattering at 8 TeV%
\thanks{Presented at ``Diffraction and Low-$x$ 2022'', Corigliano Calabro (Italy), September 24-30,
2022.}%
 }
\author{
Tamás Csörgő$^{1,~2,~\dag}$, 
Tamás Novák$^{2}$, 
\address{
$^1$Wigner RCP, H-1525 Budapest 114, POB 49, Hungary; \leavevmode\\
$^2$MATE Institute of Technology,  KRC, 
H-3200 Gy\"ongy\"os, M\'atrai \'ut 36, Hungary; 
}
\leavevmode \null \\  
Roman Pasechnik$^{3}$,
\address{
$^3$Department of Physics, Lund University, S - 223 63 Lund, Sweden;
\leavevmode \null \\  
}
\leavevmode \null \\  
András Ster$^{1}$ and István Szanyi$^{1,~2,~4}$ 
\leavevmode \\
\address{
$^4$E\"otv\"os University, H - 1117 Budapest, P\'azm\'any P. s. 1/A, Hungary;\\ \leavevmode \null \\
$^\dag${tcsorgo@cern.ch}
}
}
\maketitle
\begin{abstract}
We complete the model-independent analysis of
the scaling properties of the differential cross section of elastic proton-proton cross sections, including new TOTEM data published in 2022 at    $\sqrt{s} = 8$ TeV. 
We separate the signal and the background region with a new gating method. In the signal region, we
find that the statistical significance of Odderon exchange from the combined 7.0 and 8.0 TeV $pp$ data of TOTEM and the 1.96 TeV $p\bar{p}$ data of D0 is at least 7.32$\sigma$. In the background region, the scaling functions of elastic proton-proton data at  7 and 8 TeV and that of elastic proton-antiproton scattering data at 1.96 TeV agree with a statistical significance not larger than 1.93$\sigma$. 
\end{abstract}
  
\section{Introduction}

The Odderon by now is an almost a 50 years old scientific puzzle. 
The possible existence of the Odderon exchange was proposed by  Lukaszuk and Nicolescu in 1973 \cite{Lukaszuk:1973nt}, but until 2021,  publications of a statistically significant, at least 5$\sigma$ level observational evidence of the theoretically predicted Odderon exchange from experimental data were lacking.
Recent data from the TOTEM Collaboration at 2.76 TeV, 7 TeV, 8 TeV and 13 TeV~\cite{Antchev:2013iaa,TOTEM:2018psk,Antchev:2018edk} at the Large Hadron Collider (LHC) allowed for the discovery of the Odderon exchange, when combined with data of the D0 experiment at 1.96 TeV ~\cite{Abazov:2012qb} at Tevatron. 
Both the Tevatron and the LHC provided  collisions at the TeV energy scale, where the dominant exchanges are the gluonic Pomeron and Odderon exchanges, while the contribution of hadronic Reggeon exchanges are suppressed below the experimental errors~\cite{Broniowski:2018xbg}. 
Consequently the Odderon exchange has been searched for, as an odd component of elastic scattering amplitude which changes sign under crossing~\cite{Ster:2015esa,Csorgo:2019ewn,Csorgo:2020wmw}. 

\begin{figure}[!hbt]
 \centerline{
\includegraphics[width=0.48\textwidth]{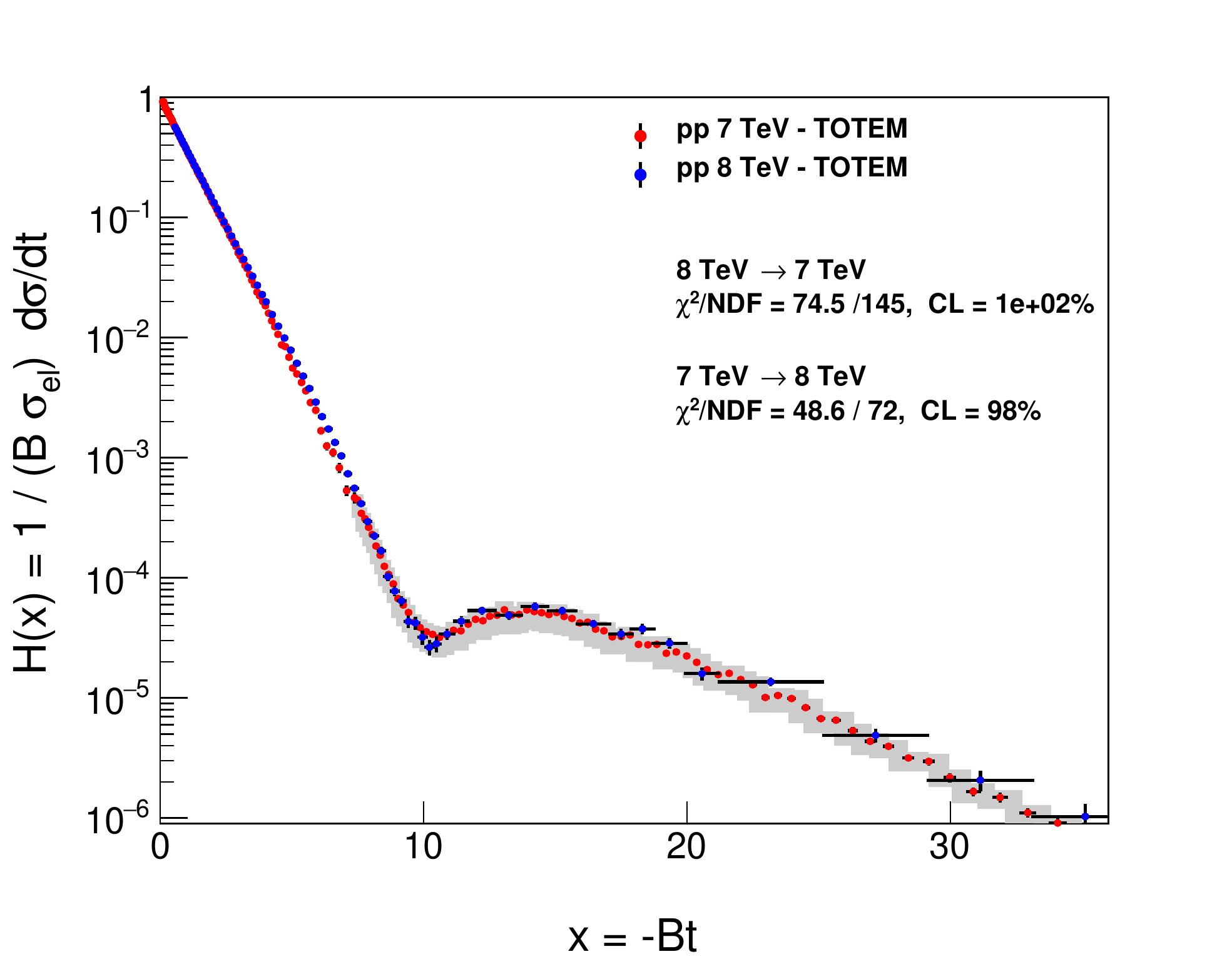}
\includegraphics[width=0.48\textwidth]{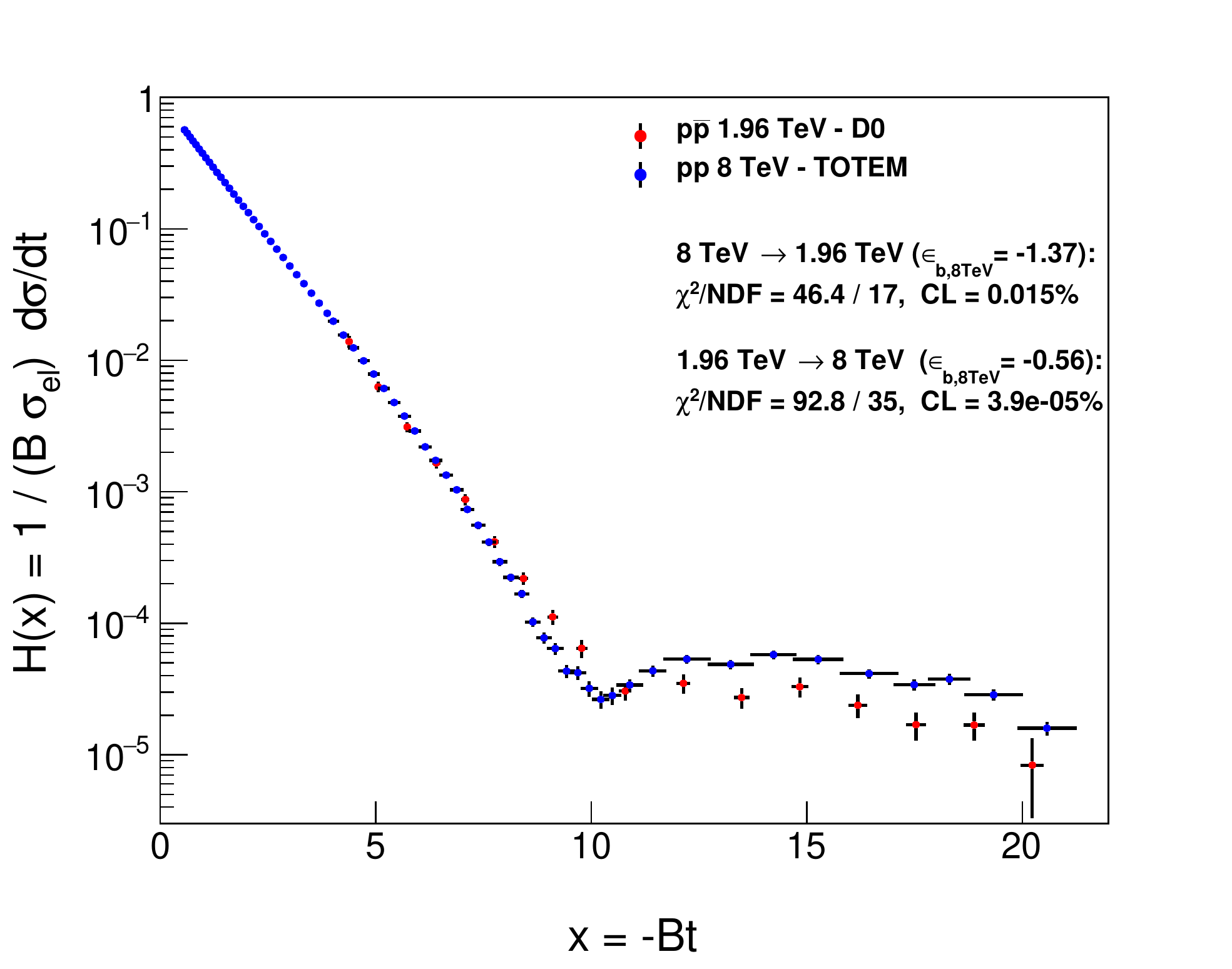}
 }
\caption{ 
Left panel 
compares the $H(x,s_i|pp)$ scaling functions at  $\sqrt{s_1} = $ 7 TeV~\cite{TOTEM:2013lle,Csorgo:2019ewn} and $\sqrt{s_2} = $ 8 TeV~\cite{TOTEM:2021imi}, corresponding to an agreement
with a confidence level of CL $\geq$ 98\%.   
Right panel 
compares the $H(x,s_2|pp)$ with the $H(x,s_3|p\bar{p})$ scaling functions  at $\sqrt{s_3} = $ 1.96 TeV~\cite{Abazov:2012qb,Csorgo:2019ewn}, corresponding to
a disagreement, indicating an Odderon signal of at least 3.74$\sigma$, when all the 17 D0 $p\bar{p}$ datapoints at  $\sqrt{s_3} = $ 1.96 TeV are compared  with $pp$ data in the same $x = - tB $ range at $\sqrt{s_2} = $ 8 TeV.
}
\label{fig:H(x)-signal-1.96-vs-8-TeV}
\end{figure}

Currently there are at least four published papers~\cite{Csorgo:2019ewn,Csorgo:2020wmw,TOTEM:2020zzr,Szanyi:2022ezh} that present statistically significant observations of the Odderon. A model-independent  and data-driven statistical method that utilized the  $H(x)$ scaling property of $pp$ elastic scattering was published in February 2021~\cite{Csorgo:2019ewn}. This method is based on a direct data-to-data comparison, without theoretical inputs, hence it is a truly model-independent method, however, the domain of the validity of this method  has been determined so far only model-dependently ~\cite{Csorgo:2019ewn}. The resulting significance of Odderon-exchange was found to be at least 6.26$\sigma$~\cite{Csorgo:2019ewn}, which 
seems to be the first published, greater than 5$\sigma$ observation of Odderon-exchange. 
 ~This discovery paper~\cite{Csorgo:2019ewn} was based on a re-analysis of previously published, public-domain experimental data of the D0 and TOTEM collaborations,
Refs.~\cite{Abazov:2012qb,TOTEM:2018psk,TOTEM:2013lle} measured at $\sqrt{s} =$ 1.96, 2.76, and 7.0 TeV, respectively. We follow here the conventions, definitions, notations and methods of Ref.~\cite{Csorgo:2019ewn},  extending it to the analysis of new  data on elastic  $pp$ collisions  at 8.0 TeV~\cite{TOTEM:2021imi},  as  illustrated in Fig.~\ref{fig:H(x)-signal-1.96-vs-8-TeV}.

The second proof of a discovery level Odderon exchange   was published in July 2021, based on extrapolations of both $pp$ and  $p\bar p$ differential cross-sections with the help of the Real Extended Bialas-Bzdak model~\cite{Bialas:2006qf,Nemes:2015iia},  leading to a significance of at least 7.08$\sigma$~\cite{Csorgo:2020wmw}. In addition to the above mentioned datasets, this analysis also utilized the $p\bar{p}$ elastic scattering data at $\sqrt{s} = 0.546$ TeV, measured by the UA4 experiment at the SPS collider~\cite{UA4:1985oqn}. 

The third observation of Odderon exchange was the result of a joint analysis of the D0 and TOTEM experimental collaborations, leading to a statistical significance of at least 5.2$\sigma$, as published in August 2021~\cite{TOTEM:2020zzr}. In contrast to the publications of Refs.~\cite{Csorgo:2019ewn,Csorgo:2020wmw}, that were based on a re-analysis of previously published experimental data already in the public domain, the D0-TOTEM paper ~\cite{TOTEM:2020zzr} was based on new experimental data. 
Another difference was that the D0-TOTEM paper ~\cite{TOTEM:2020zzr} was limited to the diffractive interference region, utilizing  8 D0 points out of all the publicly available 17 D0 datapoints, while the ReBB model of Ref.~\cite{Csorgo:2020wmw} included 14 out of the 17 D0 points in the domain of validity of its model-dependent Odderon search. In contrast, the model-independent scaling study of   Ref.~\cite{Csorgo:2019ewn} utilized all the 17 published D0 points. 

In March 2022, the TOTEM Collaboration published its 8 TeV data on elastic   $pp$ differential cross-section~\cite{TOTEM:2021imi} in an extended kinematic range. These data  have  been used in an updated Real Exendent Bialas-Bzdak model, to find that the statistical significance of the observation of Odderon exchange is so large, that 
it amounts practically to a certainty~\cite{Szanyi:2022ezh}. 

\section{$H(x|pp)$ scaling and Odderon exchange at 8 TeV}\label{sec:scaling}
In this section, we demonstrate, that
for elastic  $pp$ collisions, $H(x,s|pp) = H(x|pp) $ becomes independent of $s$, the square of the centre-of-mass energy in a limited, but large enough energy region, that includes 1.96 TeV, 2.76 TeV, 7 TeV and 8 TeV. 
Table ~\ref{tab:H(x)-scaling-pp-7-8-2.76} summarizes the results of the pairwise
comparison of the $H(x,s_i|pp)$ scaling functions for $s_i = $ 2.76 TeV, 7 TeV and 8 TeV. These datasets pairwise agree at a confidence level (CL) of at least 98\% .

\begin{table*}[!hbt]
    \centering
    \begin{tabular}{cc}
        $\sqrt{s}$ (TeV) & signal (CL, \%)  \\ \hline \hline
        8.0 vs 7.0 &  $\geq$ 98  \\ 
        8.0 vs 2.76 & $\approx$ 100 \\ 
        7.0 vs 2.76 & $\approx$ 100 \\ \hline\hline
    \end{tabular}
    \caption{ Pairwise comparision of the $H(x,s_i|pp)$ scaling functions for $\sqrt{s_i} =$ 8.0, 7.0 and 2.76 TeV indicate, that these scaling functions are energy independent with a confidence level of at least 98\%.
  }\label{tab:H(x)-scaling-pp-7-8-2.76}
\end{table*}

\begin{table*}[!hbt]
    \centering
    \begin{tabular}{cc}
        $\sqrt{s}$ (TeV) & Odderon signal ($\sigma$)  \\ \hline \hline
        8.0 vs 1.96 &  $\geq$ 3.74  \\ 
        7.0 vs 1.96 & $\geq$ 6.26 \\ 
        2.76 vs 1.96 & $\geq$ 0.01 \\ \hline 
        8.0 \& 7.0 vs 1.96 & $\geq$ 7.07 \\ 
        8.0 \& 7.0 \& 2.76 vs 1.96 & $\geq$ 5.77 \\ \hline \hline
    \end{tabular}
    \caption{ Pairwise comparision of the $H(x,s_i|pp)$ with the  $H(x,s_j|p\bar{p})$ scaling functions for $\sqrt{s_i} =$ 8.0, 7.0 and 2.76 TeV and $\sqrt{s_j} = 1.96$ 2.76 TeV indicate, that these scaling functions are clearly 
    different, except for the 
    comparison of the 2.76 TeV $pp$ and the 1.96 TeV $p\bar{p}$ dataset.
    The combined significances
     are well above 
     5$\sigma$.
  }\label{tab:H(x)-scaling-pp-pbarp-7-8-2.76}
\end{table*}

Table~\ref{tab:H(x)-scaling-pp-pbarp-7-8-2.76} shows the
comparison of the $H(x,s_i|pp)$ scaling functions with the  $H(x,s_j|p\bar{p})$ scaling functions for $\sqrt{s_i} =$ 8.0 TeV, 7.0 TeV and 2.76 TeV, and $\sqrt{s_j} = 1.96$  TeV.  These $pp$ and $p\bar{p}$ scaling functions are significantly different, except for the standalone comparison of the 2.76 TeV $pp$ and the 1.96 TeV $p\bar{p}$ dataset.
    The last two lines show the lower estimate for the minimal combined significances of these $pp$ data for an energy-independent $H(x |pp)$ as compared to the data on the $H(x,s_j|p\bar{p})$  function at $\sqrt{s_j} = 1.96$ TeV. The combined significances are safely above the discovery threshold of 5$\sigma$.  If the 2.76 TeV $pp$ dataset is not considered, the combined significance of the signal of Odderon exchange from the data measured at 8.0 TeV, 7.0 TeV and 1.96 TeV exceeds 7.0$\sigma$. This is achieved with full utilization of all the 17 D0 points and without any model-dependent input to this analysis. 
    
 So far, we have demonstrated, that
for elastic  $p\bar{p}$ collisions, $  H(x,s_i|pp) \neq H(x,s_j|p\bar{p})$ for 
 $\sqrt{s_i} =$ 8.0 TeV, 7.0 TeV and 2.76 TeV and $\sqrt{s_j} = 1.96$  TeV, which provides a statistically significant signal for Odderon exchange. However, in Ref.~\cite{Csorgo:2019ewn}, we utilized the Real Extended Bialas-Bzdak model to show that the $pp$ $H(x,s|pp)$ scaling function remains energy independent even at $\sqrt{s} = 1.96 $ TeV. 
 To complete  a statistically significant and model-independent 
    proof of Odderon-exchange, we now show  that $H(x, s_j|pp) = H(x |pp)$ is energy independent also at $\sqrt{s_j} = 1.96$ TeV.
    Given that there are no $pp$ data measured at 1.96 TeV, we separate the Odderon signal and background region, and show that $H(x, s_j  |p\bar{p}) = H(x |pp)$ is energy independent at 1.96 TeV -- if the comparison is limited to the background region, i.e.  outside the Odderon signal region.
    
\begin{table*}[!hbt]
    \centering
    \begin{tabular}{ccccc}
        $\sqrt{s}$ (TeV) & $n$ & $m$ & signal ($\sigma$) & background ($\sigma$)   \\ \hline \hline
        1.96 vs 7.0 & 3 & 2 & 6.33 &  1.70 \\ 
        1.96 vs 8.0 & 3 & 2 & 4.03 &  1.04  \\ 
        1.96 vs 7.0 vs 8.0 & 3 & 2 & 7.32 &  1.93  \\ \hline
        1.96 vs 8.0 & 6 & 1 & $\geq$ 4.55 &  0.13   \\ \hline \hline
    \end{tabular}
    \caption{Closing gates separate the  signal and the background region of Odderon exchange at
    $\sqrt{s_1} = 7$ TeV and $8$ TeV. The first $n$ and the last $m$ of the 17 D0 $p\bar{p}$ points are taken as background, at low $-t$ and large $-t$, respectively. The remaining
    $17 - n - m$  datapoints of D0 constitute the signal region.
  }\label{tab:H(x)-signal-background-7-8-1.96}
  \vspace{-5mm} 
\end{table*}

    We utilize the method of closing gates that is suitable to separate the signal and background regions of Odderon exchange. This method starts with the utilization of the full D0 dataset, then considers removing either the leftmost or the rightmost D0 datapoint, evaluating the expected increase in the signal in both cases. The next step is to gate out that of the leftmost or the rightmost datapoint, that results in a 
   smaller increase of the signal for Odderon exchange. This procedure is iterated step by step until the significance is maximized. The remaining $17 - n - m$ datapoints of D0 are considered the signal region and the gated out $n+m$ datapoints of D0 correspond to the background region of Odderon exchange.
 
 When comparing the $H(x,s_i|pp)$  with    the $H(x,s_j|p\bar{p})$ scaling functions, the best signal region is found at $(n,m) = (3,2)$, corresponding to a statistical significance of the Odderon signal that is increased from 6.26$\sigma$ to 6.33$\sigma$, while for the same $(n,m)$ selection the background regions at 1.96 TeV and 7 TeV agree: the significance of their difference is 1.70$\sigma$ only, less than an indication of a difference that starts at 3.0$\sigma$ in our terminology. For the same $(3,2)$ gating, the comparison of 8 TeV $pp$ and 1.96 TeV $p\bar{p}$ datasets yields a signal significance that increases from 3.55$\sigma$ to 4.03$\sigma$, while the backgrounds  between the $pp$ and the $p\bar{p}$
    scaling functions agree at 1.04$\sigma$. The combined significances for the signal and the background are shown for the same $(3,2)$ gating in the third line of Table~\ref{tab:H(x)-signal-background-7-8-1.96}, indicating that the backgrounds agree within 1.94$\sigma$, while the combined significance of Odderon exchange increases to 
    7.32$\sigma$. The last line shows, that
    the gated signal from the 8 TeV versus 1.96 TeV comparison can be further increased to 4.55$\sigma$ with the gate position at $(6,1)$, where the backgrounds also agree at 0.13$\sigma$.

Table \ref{tab:H(x)-signal-background-7-8-1.96} also indicates, that outside the signal region, the backgrounds of the $H(x,s |pp)$ scaling functions of elastic  $pp$ collisions at 7 TeV and 8 TeV agree with the same region of
elastic $p \bar{p}$ collisions at 1.96 TeV. Even for the combined dataset and for slightly different selections of the background region, these $pp$ and $p \bar{p}$ backgrounds agree within less than 2.0$\sigma$ deviations.
 This is possible only if the $H(x,s|pp) = H(x|pp)$ energy independent scaling relation includes the center of mass energy of $\sqrt{s_j} = 1.96$ TeV.

\section{Summary}\label{sec:summ}
We have completed, without any model-dependent ingredients, the proof that at the TeV energy scale, the $H(x,s_i|pp)$ scaling functions are independent of the center-of-mass colliding energy at $\sqrt{s_i}$ = 1.96, 2.76, 7.0 and 8.0 TeV. We have separated
the signal and the background regions of this scaling using the method of closing gates.
The exact location of the gating does not influence the qualitative conclusion, that in the signal region, the combined statistical significance of the Odderon exchange is at least 7.0$\sigma$, safely above the 5$\sigma$ discovery threshold. In the background region, the energy-independent $H(x|pp)$ scaling function agree with the  $H(x|p\bar p)$ function at $\sqrt{s} = 1.96$ TeV, with a less than 2$\sigma$ statistical significance of difference.

\section{Acknowledgments}\label{sec:ackn}
We thank A. Papa and his team for their hospitality and for organizing an inspiring and useful meeting at Corigliano Calabro (Italy). Our research has been supported by 
the Hungarian NKFI Grants K133046 and 2020-2.2.1-ED-2021-00181 and 
the {\'U}NKP-22-3 New National Excellence Program,
by the Swedish Research Council grants no. 621-2013-4287 and 2016-05996 as well as by
the European Research Council (ERC) 
under the EU’s Horizon 2020 research and innovation programme (Agreement no. 668679).

\bibliographystyle{unsrt}  
\bibliography{Odderon-DL22-APPB}

\end{document}